\documentclass[graybox]{svmult}

\usepackage{mathptmx}       
\usepackage{helvet}         
\usepackage{courier}        
\usepackage{type1cm}        
\usepackage{makeidx}         
\usepackage{graphicx}        
\usepackage{multicol}        
\usepackage[bottom]{footmisc}
\usepackage[numbers]{natbib}

\newcommand{\pd}{$P_\delta$}
\newcommand{\pln}{$P_{\ln(1+\delta)}$}

\begin{document}

\title*{Gaussianization: Enhancing the Statistical Power of the Power Spectrum}
\author{Mark C.\ Neyrinck}
\institute{Mark C.\ Neyrinck \at The Johns Hopkins University, Baltimore, MD 21218}
%
%
\maketitle

\abstract{The power spectrum is widely used in astronomy, to analyze
  temporal or spatial structure.  In cosmology, it is used to quantify
  large-scale structure (LSS) and the cosmic microwave background
  (CMB).  This is because the power spectrum completely quantifies
  Gaussian random fields, which the CMB and LSS fields seem to be at
  early epochs.  However, at late epochs and small scales,
  cosmological density fields become highly non-Gaussian.  The power
  spectrum loses power to describe LSS and CMB fields on small scales,
  most obviously through high covariance in the power spectrum as a
  function of scale.  Practically, this significantly degrades
  constraints that observations can place on cosmological parameters.
  However, if a nonlinear transformation that produces a (more)
  Gaussian 1-point distribution is applied to a field, the extra
  covariance in the field's power spectrum can be dramatically
  reduced.  In the case of the roughly lognormal low-redshift matter
  density field, a log transform accomplishes this.  Applying a log
  transform to the density field before measuring the power spectrum
  also tightens cosmological parameter constraints by a factor of
  several.}

\bigskip

A Gaussian random field has convenient statistical properties.  Its
meaningful information is fully quantified by the power spectrum; all
connected higher-order statistics vanish.  Of particular importance
for a measurer of (parameters which depend on) the power spectrum, all
off-diagonal power-spectrum covariance matrix elements vanish for a
Gaussian random field.

\begin{figure}
  \begin{center}
    \includegraphics[scale=.5]{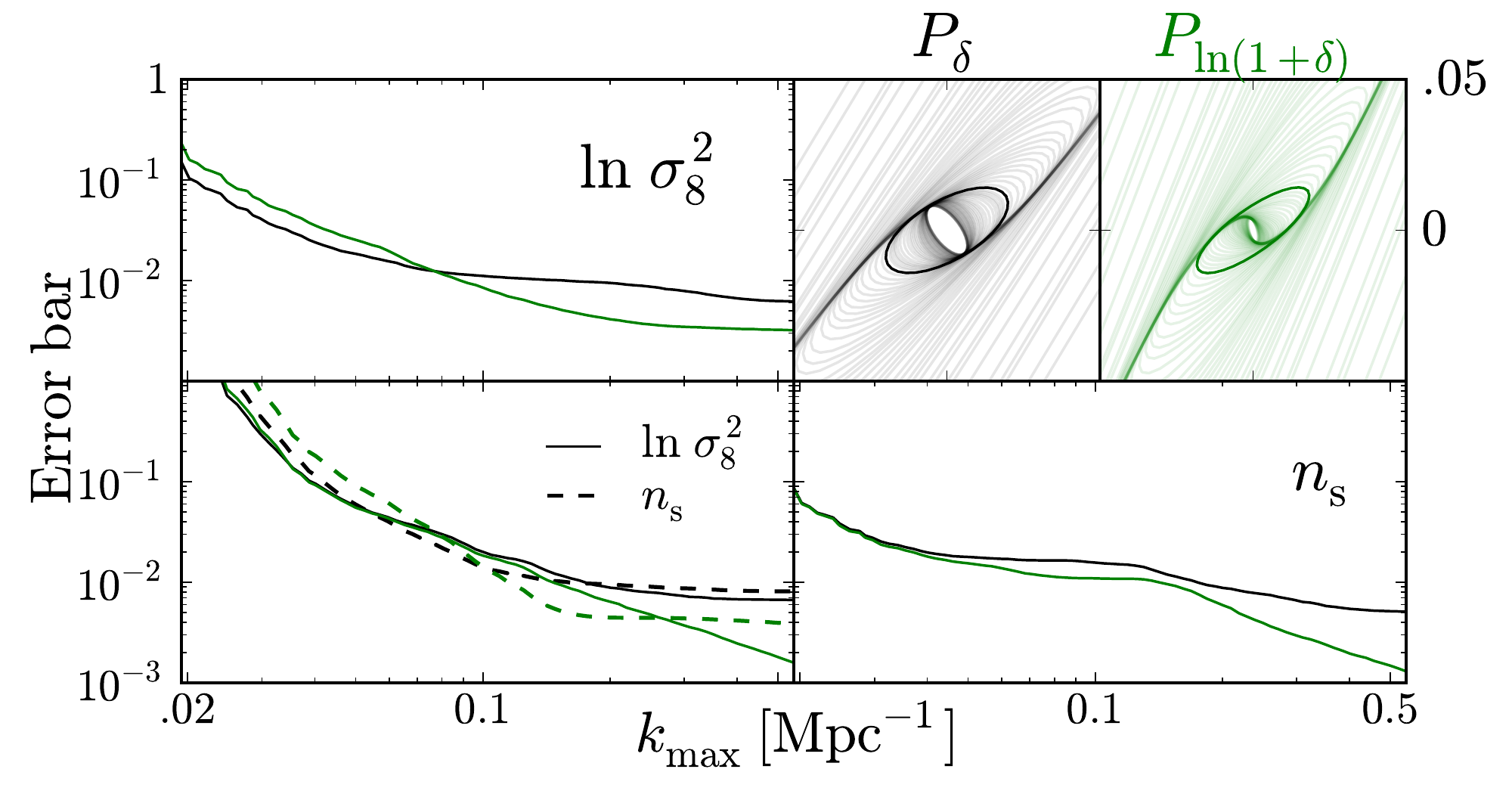}
  \end{center}
  \vspace{-0.35cm}
  \caption{Fisher-matrix estimates of error-bar (half-)widths and
    error ellipses for the cosmological parameters $\ln \sigma_8^2$
    and $n_s$, inferred analyzing \pd\ (black) and \pln\ (green).  We
    show how they depend on the maximum wavenumber $k_{\rm max}$
    included in a power-spectrum analysis of a 1-Gpc real-space matter
    density field.  Diagonal panels show unmarginalized error bars
    over single parameters.  In the lower-left panel, errors in each
    parameter are marginalized over the other.  The upper-right panel
    shows how error ellipses contract as $k_{\rm max}$ increases.
    There is an ellipse shown for each $k_{\rm max}$ constituting the
    curves in the other panels.  Outside the bold ellipses, analyzing
    only large scales where $k_{\rm max} < 0.1$\,Mpc$^{-1}$, \pd\ and
    \pln\ perform similarly.  Inside the bold ellipses, nonlinear
    scales are included, up to the innermost ellipse that corresponds
    to $k_{\rm max}=0.5$ Mpc$^{-1}$.  Here, \pln\ greatly outperforms
    \pd.  See Ref.\ \cite{n11b} for more details.  }
  \label{fig:ellipses_s8n}
\end{figure}

On small scales at late epochs, the cosmological overdensity field
$\delta=\frac{\rho}{\langle\rho\rangle}-1$ is highly non-Gaussian,
with high off-diagonal power-spectrum covariance.  However, this
non-Gaussianity is not of an essential form, and can be largely
removed with a monotonic transformation.  The
near-lognormality \cite{colesjones} of the density field may be
exploited; a Gaussianizing transform like $\delta\to\ln(1+\delta)$
much reduces the power-spectrum covariance \cite{nss09}.  As shown in
Fig.\ \ref{fig:ellipses_s8n}, it also reduces error bars on inferred
parameters, reaching a factor of 5 reduction in the best case of the
tilt $n_s$.

\begin{figure}
  \begin{center}
    \includegraphics[scale=.5]{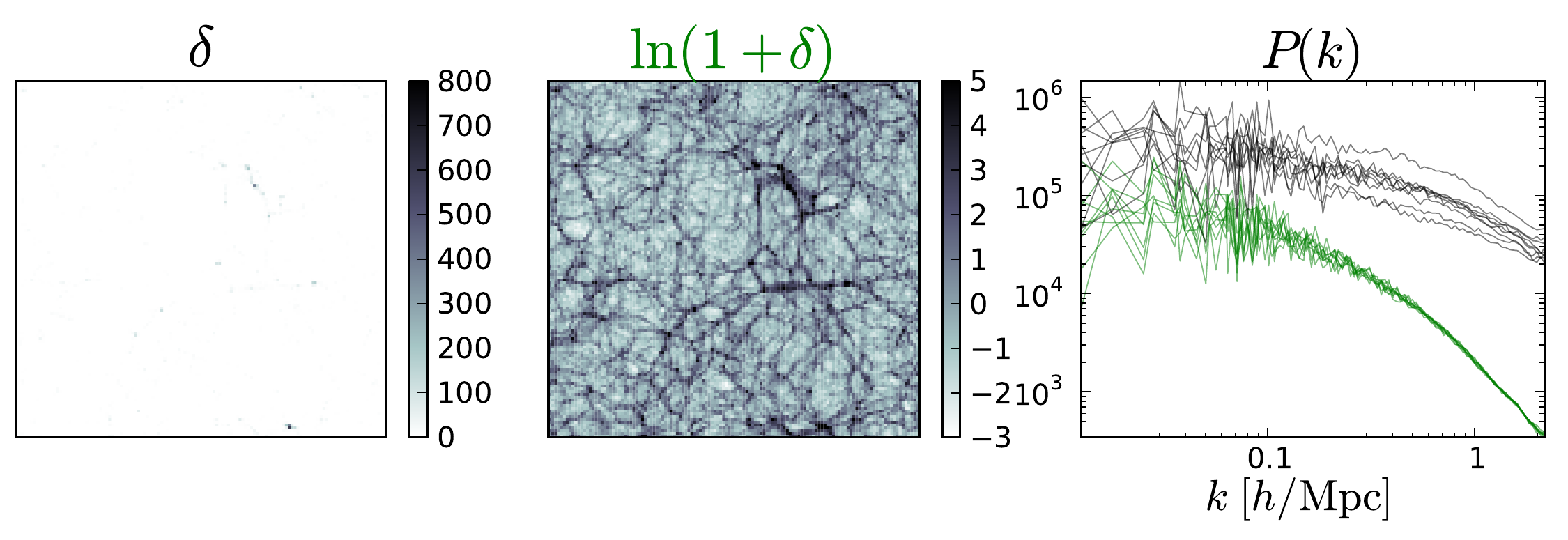}
  \end{center}
  \vspace{-0.5cm}
  \caption{Left: $\delta$ in a 2-$h^{-1}$\, Mpc slice of the
    500-$h^{-1}$\, Mpc Millennium simulation (MS), viewed with an
    unfortunate linear color scale.  Middle: the same slice with a
    logarithmic color scale.  Right: the 2D power spectra \pd\ and
    \pln\ of $\delta$ (black) and $\ln(1+\delta)$ (green), in 9 such
    slices.  The wild, coherent fluctuations in \pd\ illustrate its
    high (co)variance, absent in \pln.  }
  \label{fig:pows}
\end{figure}
\vspace{-1cm}

\bibliographystyle{spphys}

\end{document}